# Mechanical detection and imaging of hyperbolic phonon polaritons in hexagonal Boron Nitride

Antonio Ambrosio[1,2,3], Luis A. Jauregui[2], Siyan Dai[4], Kundan Chaudhary[5], Michele Tamagnone[5], Michael Fogler[4], Dimitri N. Basov[4,6], Federico Capasso[5], Philip Kim[2] and William L. Wilson[1]

[1]*Center for Nanoscale Systems, Harvard University, Cambridge, Massachusetts 02138, USA*

[2]*Department of Physics, Harvard University, Cambridge, Massachusetts 02138, USA*

[3]*CNR-SPIN U.O.S. Napoli, Complesso Universitario di Monte Sant'Angelo, Via Cintia, 80126 – Napoli, Italy*

[4]*Department of Physics University of California, San Diego (UCSD)*

[5]*Harvard John A. Paulson School of Engineering and Applied Sciences, Harvard University, Cambridge, Massachusetts 02138, USA*

[6]*Department of Physics, Columbia University, 538 West 120th Street, New York, New York 10027, United States*

corresponding author: *ambrosio@seas.harvard.edu*

**Abstract** – Mid-infrared nano-imaging and spectroscopy of two-dimensional (2D) materials have been limited so far to scattering-type Scanning Near-field Optical Microscopy (s-NSOM) experiments where light from the sample is scattered by a metallic-coated Atomic Force Microscope (AFM) tip interacting with the material at the nanoscale. These experiments have recently allowed imaging of plasmon polaritons in graphene as well as hyperbolic phonon polaritons ($HP^2$) in hexagonal Boron Nitride (hBN). Here we show that the high mechanical sensitivity of an AFM cantilever can be exploited for imaging hyperbolic phonon polaritons in hBN. In our imaging process, the lattice vibrations of hBN micrometer-sized flakes are locally enhanced by the launched phonon polaritons. These enhanced vibrations are coupled to the AFM tip in contact to the sample surface and recorded during scanning. Imaging resolution better than λ/20 is showed, comparable to the best resolution in s-NSOM. Importantly, this detection mechanism is free from light background and it is in fact the first photon-less detection of phonon polaritons.

The first Atomic Force Microscope (AFM) was evocatively labeled "Touching Microscope" to convey that the microscope "feels" the sample's atoms and can produce an atomically resolved image of the surface. In their first paper "Atomic Force Microscope" [1], Binning, Quate and Gerber had already envisioned a general-purpose device "that will measure any type of force; not only interatomic forces, but electromagnetic forces as well". Today, the AFM is really a general-purpose device used in many configurations for specific force characterization, including Electric Force Microscopy [2,3], Magnetic Force Microscopy [4,5], Microwave Impedance Microscopy [6,7], Multi-Frequency Force microscopy [8], etc. In its most recent applications AFM has also been proposed in different schemes for optical near-field imaging [9,10,11,12,13] and spectroscopy at the nanoscale without detecting any light [14,15,16,17,18,19].

In this paper, we show imaging of optically excited hyperbolic phonon polaritons ($HP^2$) in hexagonal Boron Nitride (hBN) flakes by monitoring only the mechanical oscillations induced in an AFM cantilever.

Phonon Polaritons, as well as Surface Plasmon Polaritons (SPPs) have attracted large interest for many years due to their role as energy carriers [20,21,22]. In fact, directional control of electromagnetic energy propagation in flat optoelectronic devices needs strong coupling of electromagnetic waves to local material excitations. SPPs are generated by collective excitations of electrons (e.g. at a metal-dielectric interface), while an electromagnetic wave coupled to the lattice vibrations (phonons) of a polar crystal gives rise to the excitation of phonon polaritons.

Hexagonal boron nitride is of particular interest in terms of phonon polaritons since it has been found to have low loss phonon polaritons in the upper Reststrahlen band (1370-1610cm$^{-1}$). Moreover, in this wavelengths (wavenumbers) range (Type II band), hBN is hyperbolic, i.e. $\varepsilon_z > 0$ and $\varepsilon_t < 0$, where $\varepsilon_z$ and $\varepsilon_t$ are the axial (normal to the surface) and tangential (in plane) permittivities, respectively [23,24,25,26,27]. As a consequence, the polariton beam propagates inside hBN along specific directions at an angle $\theta$, provided by the relation $tan[\theta] = i\ (\varepsilon_t/\ \varepsilon_z)^{1/2}$ from the point of origin [28,29]. In fact, due to the momentum mismatch between the free space illuminating light and the highly-confined propagating polaritons, a local scatterer (point of origin) close to the sample surface is needed to launch phonon polariton propagating waves [30].

Hyperbolic phonon polaritons (HP$^2$) have been previously imaged in hBN by means of scattering-type Scanning Near-Field Optical Microscopy (s-NSOM) [31,32]. In this configuration, as reported in ref. [31] an hBN flake is usually shaped in a triangle by electron beam lithography and reactive ion etching (Supporting Information). The local scatterer that allows overcoming the momentum mismatch is the metal-coated tip of the s-NSOM microscope illuminated by mid-IR light at an angle [33,34,35]. The presence of sharp edges like in hBN flakes with a well-defined geometric shape of several microns is important since hyperbolic phonon polaritons have low losses and, by reflecting at the edges, they create standing waves that are then visualized during the imaging process. Described more explicitly: 1) the tip launches the phonon polariton; 2) the polariton propagates and it is reflected at the edges of the flake creating a standing wave that may have a minimum, a maximum or any intermediate amplitude at the tip position; 3) the tip scatters the radiation to open air for far-field detection by means of a mid-IR detector. Figure 1 (a) and (b) show the amplitude and phase of the near-field scattered light (s-NSOM signal) from the surface of a 67nm thick hBN flake on 285nm thick SiO$_2$. The phonon polariton standing wave is clearly visible. The s-NSOM microscope used here is the same as that in reference [31] (Neaspec GmbH).

The detection scheme presented here is still based on an Atomic Force Microscopy platform but no radiation is detected. Imaging of the $HP^2$ is obtained by only monitoring the mechanical oscillations of the microscope cantilever (Figure 1 (c)). This detection is then free from light background. The detection scheme is similar to both Photo-thermal Microscopy and Photo-induced Force Microscopy where a light-driven local surface-tip interaction (local heating or optical-forces) is monitored as induced mechanical oscillations on an AFM cantilever [17].

In our microscope (Anasys Instruments) an infrared laser beam from an Optical Parametric Oscillator (OPO) is focused onto the tip-sample region by means of a low numerical aperture parabolic mirror (Figure 1 (c)). The laser is pulsed (10ns pulse width, 1kHz fixed repetition rate) and can be wavelength tuned in the upper Reststrahlen band of hBN (1370 - 1610cm$^{-1}$). The microscope operates as an AFM in contact mode. That is, the tip is then not oscillated by an external dithering piezo, rather it is in "contact" with the sample surface. When the sample is illuminated, some oscillations on the microscope cantilever appear at around 170kHz (one of the mechanical modes of the cantilever). Figure 1 (d) shows the image ("photo-thermal" signal) that we obtained for the polariton standing wave in the hBN flake. This image is then not obtained by detecting scattered light but by monitoring the amplitude of the induced oscillations on the cantilever at 170kHZ (with a bandwidth of 50kHz). There is complete superposition between the s-NSOM patterns shown in Figure 1 (a) and (b) and that in Figure 1 (d) (see also Figure S1). Figure 1 (e) shows instead a zoomed-in image of the polariton pattern in a corner of the shaped flake. This figure clearly shows the possibility to visualize features only a couple of hundreds nanometers wide (better than λ/20), comparable to the best resolution achieved in mid-IR s-NSOM.

In our imaging process, the lattice vibrations are locally enhanced by the launched phonon polariton, spatially modulated by the polariton standing wave. These enhanced vibrations (on a macroscopic scale this corresponds to local heating and thermal expansion) are what in fact cause the induced tip oscillations (photo-thermal signal). More precisely, the repetition rate of the illuminating laser is 1kHz. This results in periodic oscillations of the sample surface upon heating that are transferred to the microscope probe cantilever. The induced oscillations couple to several modes of the cantilever, including the mode around 170kHz whose amplitude is monitored during the imaging process.

Note that other phenomena can in principle contribute to the induced mechanical oscillations of the probe cantilever.
An enhanced radiative heat transfer, mediated by phonon-polaritons, between two objects have been for instance studied [36,37]. However, those experiments need to be carried under vacuum to get reasonable signal to noise ratio in the detection. We operated in air. Moreover, the largest enhancement of those experiments has been reported between similar polar crystals that support phonon polaritons at the same wavelength (near-field interaction of the polaritonic field). In our experiment instead, one of the objects, the microscope cantilever, is gold coated. For these reasons, it looks unrealistic that radiative heat-transfer it is our main sensing mechanism.
Direct forces of optical origin are also possible in principle, although extremely weak. According to this picture, the evanescent tail of the phonon polariton in air would couple to the metallic tip generating a weak optical force depending on the polariton local strength. This would be similar to the working principle of the Photo-induced

Force Microscopy technique [38]. However, in our experiment local heating of the sample surface is evident when the laser power approach 1mW (Supporting Information). For this, we also exclude optical forces at the sample surfaces as the detection mechanism in our experiment.

In a complementary experiment we used a hBN flake, 140nm thick, deposited on Si substrate and cut in the shape of a disk (Figure 2). In this case we changed the wavelength of the illuminating light observing how the polariton pattern periodicity changes (polariton dispersion): smaller wavenumbers result in lager fringe spacing. The periodicity of the fringes has a linear trend with the wavenumber of the illuminating light (Figure S2), in agreement with previous s-NSOM data in the upper Reststrahlen band [31]. As shown in Figure 2 (b), when the microscope probe is on top of the hBN flake, the amplitude of the photo-thermal signal as a function of the laser wavenumber reflect the Type II band with a maximum around 1368cm$^{-1}$ [31]. However, if the tip is right at the position where one of the bright fringes of the HP$^2$ standing wave is located for a specific wavelength, the photo-thermal signal recorded as a function of the illuminating wavenumber (photo-thermal spectrum) may show a peak at the wavenumber that produce the specific fringe. This case is illustrated in Figure 3 (c) where the photo-thermal spectrum shows a peak at 1440cm$^{-1}$. In this case the microscope tip is positioned on top of one of the bright fringes produced on the surface when illuminating at 1440cm$^{-1}$ (Figure 3 (b) red circle).

Some of our samples of hBN flakes are on 285nm SiO$_2$ that has a phonon resonance around 1100cm$^{-1}$. Our imaging process is able to detect this resonance as well. Figure 3 (d) shows the photo-thermal spectrum when the microscope tip is on the SiO$_2$ substrate. High signal around the SiO$_2$ phonon peak is evident.

In conclusion, in this work we show that the interference patterns created by phonon polaritons in shaped flakes of hexagonal Boron Nitride can be imaged with high-resolution by monitoring the mechanical oscillations of an AFM cantilever in contact with the sample surface. We show that this imaging process is based on local sample heating mediated by the phonon polariton resonances of hBN. This scheme is similar to photo-thermal microscopy that has shown the best results so far mainly on soft and easily deformable materials like polymers [ 39 ]. Our imaging process provides the same features obtained in scattering near-field optical microscopy but does not require detecting any light and as such is a light-background-free photon-less detection of the coupling of light to the crystal lattice oscillations. This result extends mid-IR nano-imaging and spectroscopy via mechanical detection based on an AFM probe to crystals, 2D materials and van der Waals heterostructures.

**Acknowledgment**
This work was performed in part at the Center for Nanoscale Systems (CNS), a member of the National Nanotechnology Coordinated Infrastructure (NNCI), which is supported by the National Science Foundation under NSF award no. 1541959. CNS is a part of Harvard University. SND, MF and DNB are supported by ONR N00014-15-1-2671 . D.N.B. is the Moore Investigator in Quantum Materials EPIQS program GBMF4533. This work was supported by the National Science Foundation EFRI 2-DARE program through Grant No. 1542807.

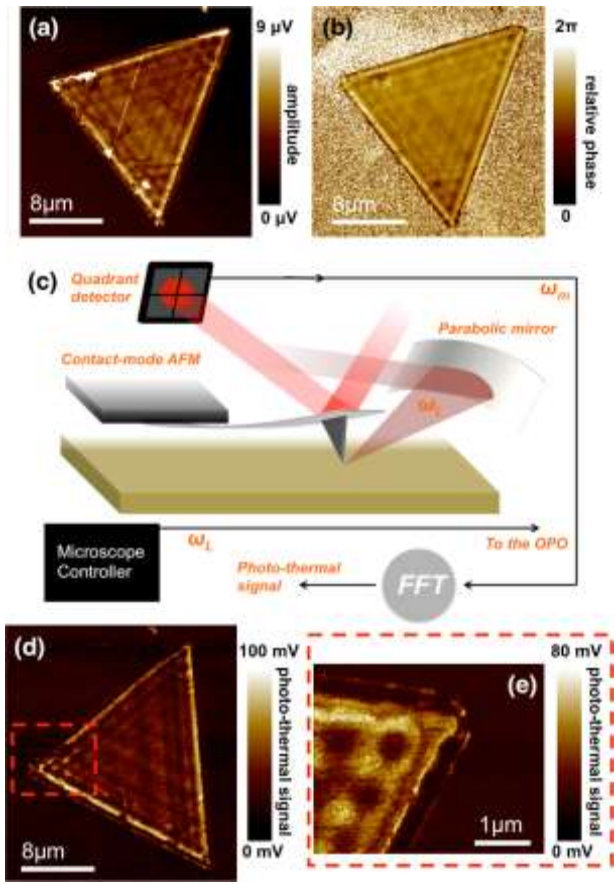

Figure 1 – (a) Amplitude (in Volts from the detector) and (b) Phase (relative phase with respect to a gold coated surface used as reference) of hyperbolic phonon polaritons imaged using s-NSOM. The hBN flake was shaped in a triangle of dimensions 16μm×16μm×18μm using electron beam lithography (Supporting Information) and is on 285nm thick $SiO_2$. The flake thickness is 67nm, measured with atomic force microscopy in contact mode. (c) Schematic of the experimental setup for photo-thermal microscopy. A pulsed laser with repetition rate $\omega_L$ = 1kHz (a mid-IR OPO laser) is focused to the tip-sample region by means of a parabolic mirror. In this imaging process, the microscope is operated with a gold-coated tip in contact with the sample surface. The tip launches the phonon polariton; the polariton propagates and is reflected at the edges of the flake creating a standing wave; the light-induced oscillations at the sample surface are transferred to the tip that start oscillating on its mechanical modes. Fourier transform of the tip oscillations, monitored through the quadrant detector, is performed to analyze the spectral content. The oscillation amplitude (in Volts from the quadrant photo-detector) at a specific frequency $\omega_m$ is then chosen (with a bandwidth of about 50kHz) to be monitored and recorded. This frequency is usually around 170kHz for the AFM cantilever we used. The recorded oscillation amplitude at $\omega_m$ is the "photo-thermal" signal used to generate figures (d) and (e). (d) Photo-thermal image of the hyperbolic phonon polariton pattern at the surface of the same flake of Figure 1 (a) and (b). (e) Zoomed-in photo-thermal image of one of the corners of the flake. The illuminating light has a wavenumber of 1440cm$^{-1}$.

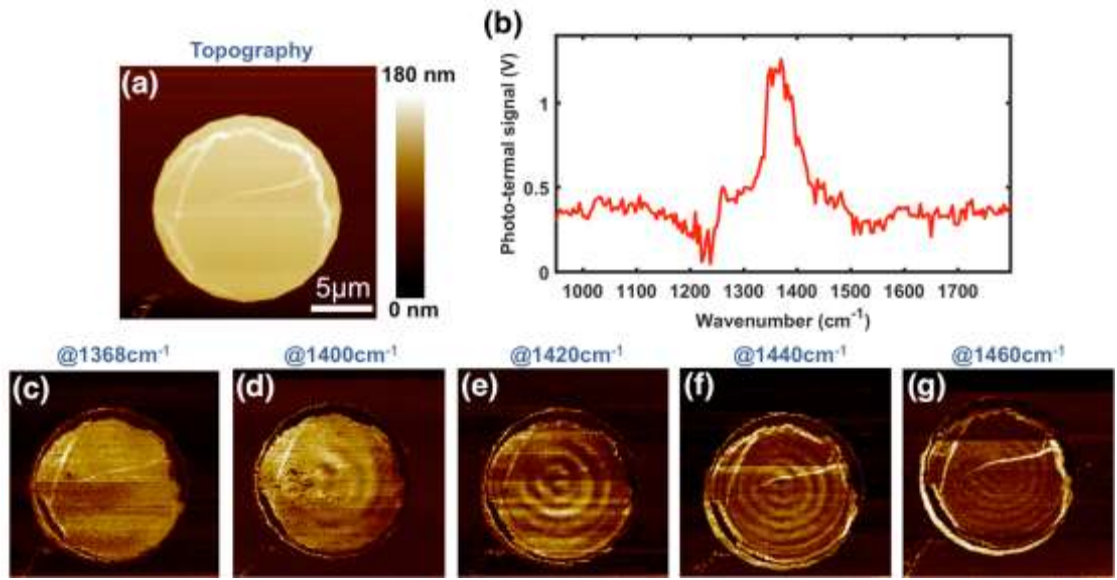

Figure 2 – (a) Topography image of an hBN flake cut in the shape of a disk. The flake is on Si substrate and is 140nm thick. The flake diameter is 15μm. (b) Photo-thermal signal as a function of the laser wavenumber. The position of the AFM tip in this case is in the central part of the flake. Signal is present in the Type II band (1370-1610cm$^{-1}$) with a peak around 1368cm$^{-1}$. (c)-(g) Photo-thermal image of the hyperbolic phonon polariton pattern at different wavelengths (wavenumbers) of the illuminating light. The periodicity of the pattern decreases by increasing the wavenumber.

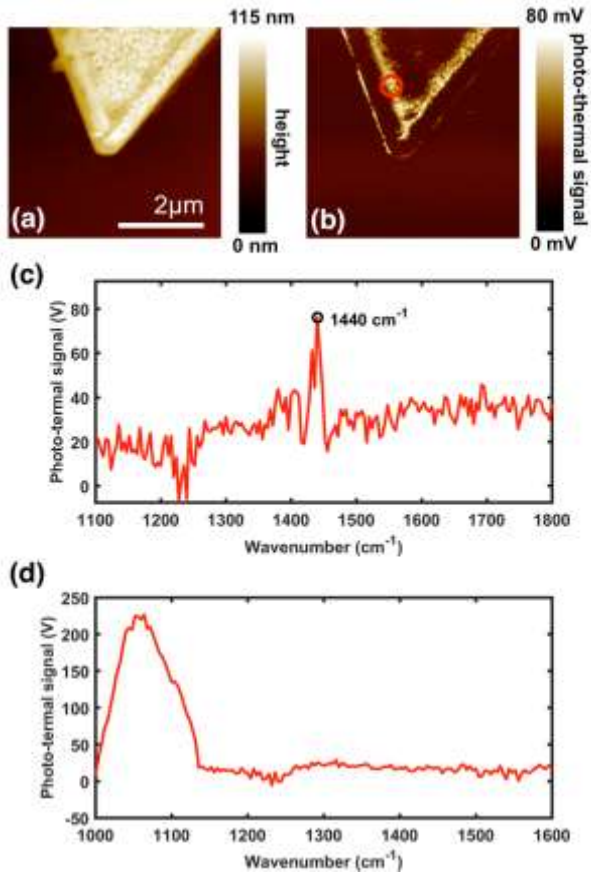

Figure 3 – (a) Topography and (b) photo-thermal image at 1440cm$^{-1}$ of a corner of a triangle-shaped hBN flake. (c) Spectrum at the tip position highlighted by the red circle in (b). This spectrum is obtained by recording the photo-thermal signal while swiping the wavenumbers of the illuminating light for a fixed position of the tip. A peak at 1440cm$^{-1}$ is evident. This only happens when the tip is on top of a maximum of interference of the polariton beam. (d) Photo-thermal spectrum recorded when the tip is on the SiO$_2$ substrate. The phonon resonance of SiO$_2$ is clearly visible around 1100cm$^{-1}$.

# Supporting Information

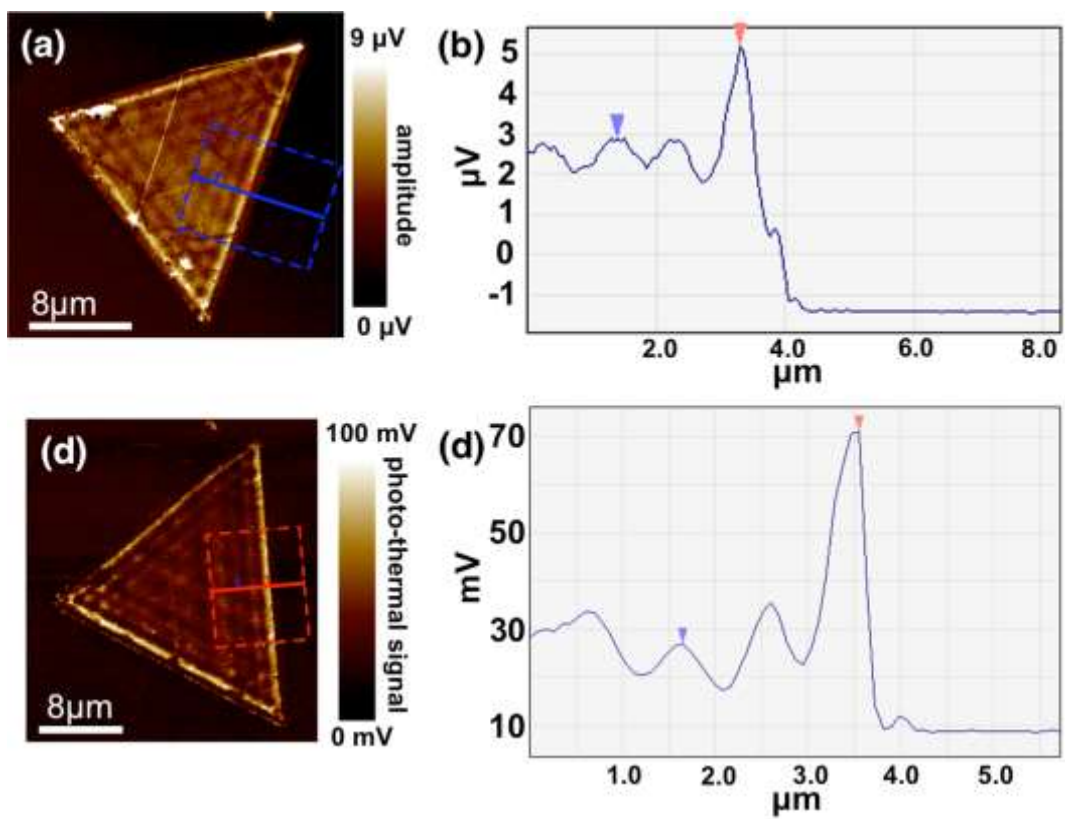

Figure S1 – (a) Scattering near-field optical image of a triangle-shaped hBN flake (Amplitude). The flake is the same of Figure 1 of the main text. (b) Profile along the blue line of Figure S1 (a) (averaged in the dashed blue contour area). (c) and (d) same as (a) and (b) for the photo-thermal image obtained by monitoring the microscope probe oscillations around 170kHz. The periodicity of the phonon polariton interference pattern (half of the distance between the blue and red arrows in the images) is 850 m for both (b) and (d).

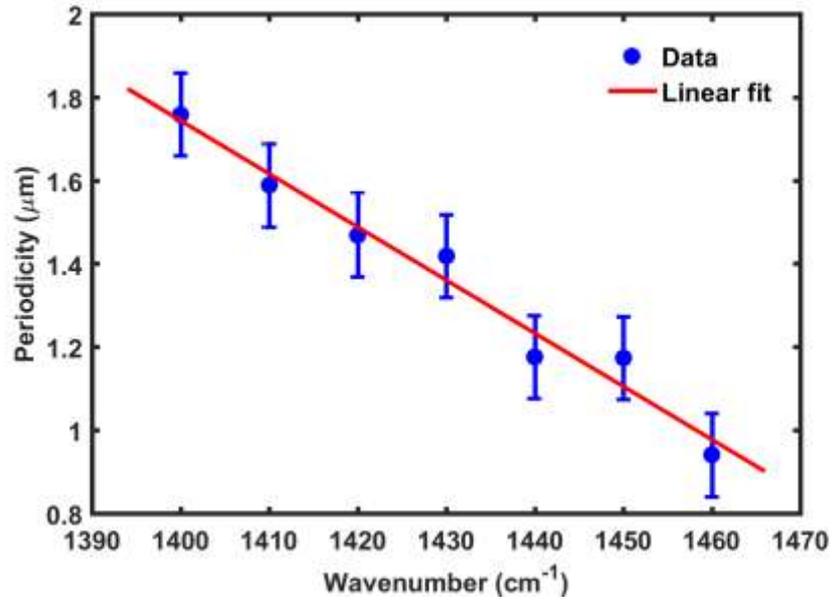

Figure S2 – Periodicity of the phonon polariton interference pattern as a function of the illuminating light wavenumber for the disk-shaped hBN flake of Figure 2. The error bars are 0.2μm.

**Artifacts in case of high laser power conditions**

When the illuminating laser power is high (about 1mW), the topography image may also show the same features of the photo-thermal image, with the topography peaks corresponding to the peaks of the induced mechanical oscillations (Figure S3). Also, the surface of the sample is usually damaged after scanning in such conditions, with permanent deformations (Figure S3). Looking at the topographical deformation during photo-thermal imaging, the variation is close to 20% of the sample thickness with an apparent reduction of the sample height (see also Figure S4). Although the sign of the thermal deformation may be compatible with the negative thermal constant components of hBN [1], such morphology variation, if true, would need a temperature variation so high to actually melt the sample. We believe, in fact, that the significant morphology deformation measured is an artifact due to our high-power operating conditions and the presence of water. In fact, the same sample after thermal annealing produces less or not at all modulation of the sample topography (Figure S5). To understand the possible artifact mechanism, it is important to consider that the tip is in contact mode, so only few nanometers far from the surface and that the contact is continuous during scanning (it is quite different from tapping mode used in scattering NSOM where the tip oscillates at about 300kHz with intermittent contact with the surface and an average distance of tens of nanometers). We believe that the local heating associated to the phonon polariton maxima may affects, besides the crystal lattice, also the water layer on the surface (water meniscus) and the intra-layer water in the flake. This can result in a local change of the interaction between the sample surface and the microscope probe that is in fact interpreted by the microscope feedback system as a local change of the sample height. Moreover, the sample height after the photo-thermal imaging in these experimental conditions (high laser power) is found to be a few nanometers higher than the original (Figure S4) while the lateral dimensions of the flake never change.

These results clearly support the thermal origin of our imaging mechanism.

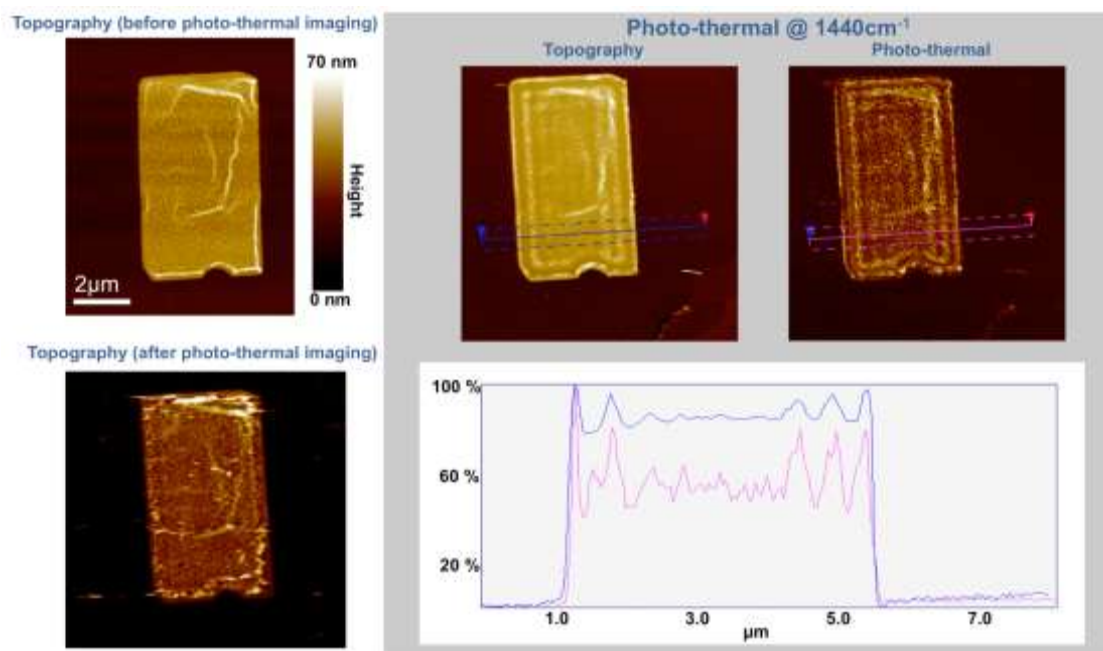

Figure S3 – (a) Topography image of a rectangle-shaped hBN flake of dimensions 4.4μm×7.6μm on 300nm thick $SiO_2$. The flake in (a) was never exposed to laser light before. (b) and (c) topography and photo-thermal image recorded during the photo-thermal imaging with a laser power up to 20% of the maximum (about 1mW). It is evident that the same features are present in both topography and photo-thermal images. This is also highlighted in the plot (d) where a cross section of the topography image is compared with the profile of the photo-thermal image. The ordinate values are reported in percent of the higher value since the purpose of this plot is only to compare the features' positions. From this, it is evident that the peaks in topography correspond to the peaks in the photo-thermal image. (e) Topography of the flake after photo-thermal imaging (no laser is sent to the microscope). The surface of the flake is now rougher compared to the pristine condition of (a) with some topographical

features that still resemble the features in (b) and (c). This is an evidence of the thermal nature of the process and the permanent alteration of the sample surface in these "high-laser-power" experimental conditions.

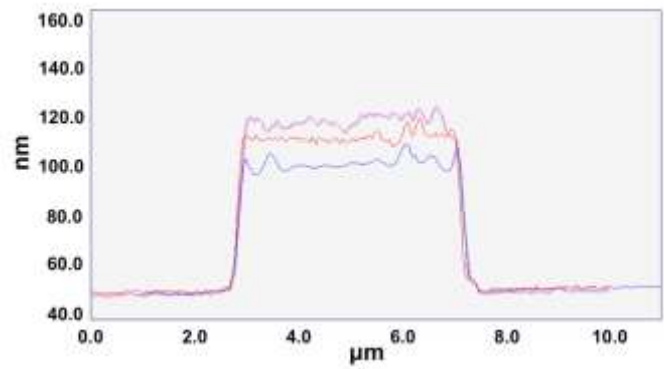

Figure S4 – Topography profiles of the flake in Figure S3. (Red) pristine conditions, sample never exposed to light. (Blue) Topography recorded during photo-thermal imaging. (Pink) Topography recorded after photo-thermal imaging.

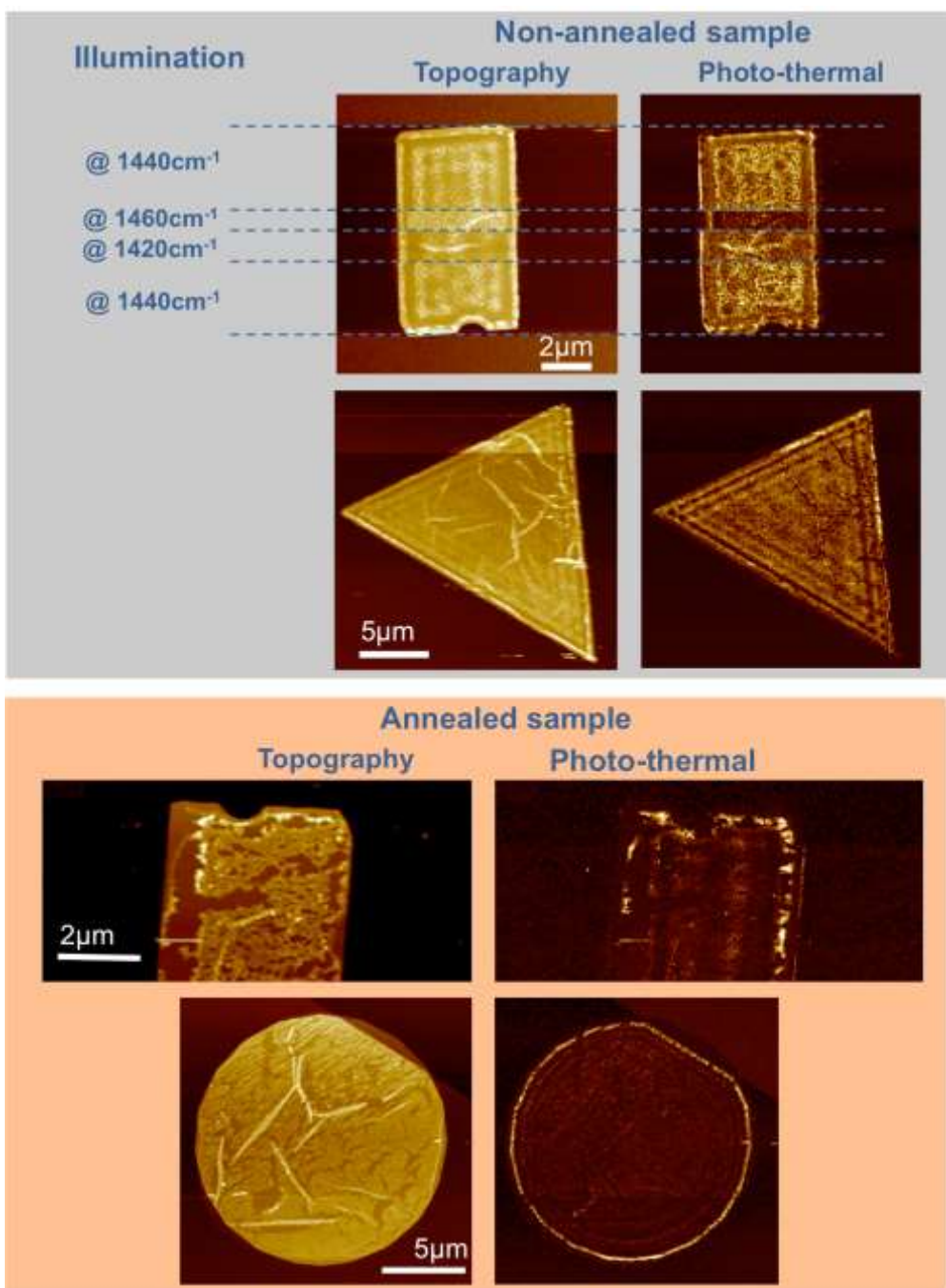

Figure S5 – Topography and photo-thermal signal for several shaped-flakes before and after the thermal annealing. The Topography deformation resembling the phonon polariton pattern is only visible for before-annealing samples. The residuals visible in topography are from the lithography process and are created during the h-BN dry etching process, which is performed using a fluorine based chemistry (see below for a full process description). During this process, the resist is also slightly etched, and,

being organic, it produces a PTFE-like fluorinated residue which is resistant to the Remover PG and chloroform and therefore remains on the sample.

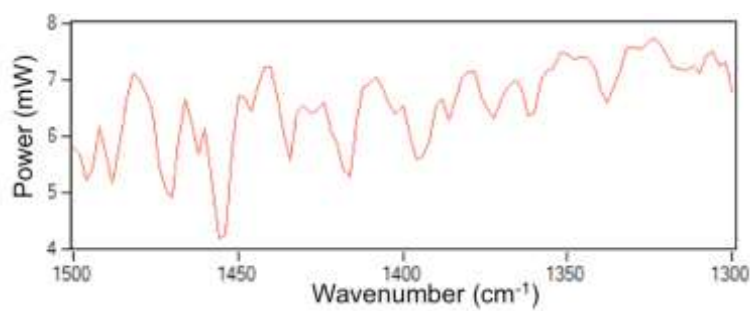

Figure S6 – Laser Power as a function of the wavelength (wavenumbers in cm$^{-1}$). This power is measured inside the microscope far from the sample. The user can decide the percentage of power illuminating the sample that represents only an upper limit of the power that is in fact illuminating the sample. The laser is focused on the sample by means of a low NA parabolic mirror at 70° from the vertical to the sample surface.

**Sample fabrication**

Hexagonal Boron Nitride (hBN) is mechanically exfoliated on to a 285nm SiO$_2$/Silicon or a bare Silicon substrate with pre-defined metallic alignment marks. The substrates are then covered with MA-N 2403 (negative e-beam resist) and exposed to an e-beam system with a dose of 1200 μC/cm$^2$ using an accelerating voltage of 125kV. The samples are shaped into triangles, circles or rectangles. After developing in AZ-726 for 1 minute, the samples are pre-baked at 100C for 10 minutes. Then hBN is etched by using a reactive ion etching system with CHF$_3$/Ar/O$_2$ at flows of 10/5/2 sccm respectively and a RF generator at 30W for 2 - 5 minutes [2]. After the etching process, exposed MA-N 2403 resist is removed by Remover PG and chloroform; afterwards, the samples are rinsed with isopropyl alcohol and dried with Nitrogen.